\documentclass{JHEP3}

\usepackage{amsmath}
\usepackage{epsfig}
\usepackage{graphicx}

\newcommand{\be}{\begin{equation}}
\newcommand{\ee}{\end{equation}}
\newcommand{\bea}{\begin{eqnarray}}
\newcommand{\eea}{\end{eqnarray}}
\newcommand{\ba}{\begin{array}}
\newcommand{\ea}{\end{array}}

\font\mybb=msbm10 at 10pt
\def\bb#1{\hbox{\mybb#1}}

\def\bI {\bb{I}}

\def\bfom{\boldmath $\omega$}

\makeatletter 
\@addtoreset{equation}{section} 
\makeatother 
 
\def\appendix#1{
  \addtocounter{section}{1}
  \setcounter{equation}{0}
  \renewcommand{\thesection}{\Alph{section}}
  \section*{Appendix \thesection\protect\indent \parbox[t]{11.15cm}
  {#1} }
  \addcontentsline{toc}{section}{Appendix \thesection\ \ \ #1}
  }

\preprint{UB-ECM-PF-09/03\\
DAMTP-2009-14}

\title{Relativistic Kinematics and Stationary Motions}

\author{Jorge G. Russo\\
Instituci\'o Catalana de Recerca i Estudis
Avan\c cats (ICREA),
\\
Departament ECM and Institut de Ciencies del Cosmos, 
\\
Facultat de F\'{\i}sica, Universitat de Barcelona,
\\
Diagonal 647, E-08028 Barcelona,
Spain.\\
}

\author{Paul K. Townsend\\
Department of Applied
Mathematics and Theoretical Physics \\
Centre for Mathematical
Sciences, University of Cambridge\\
Wilberforce Road, Cambridge, CB3
0WA,
UK.\\ 
}

\abstract{The relativistic jerk, snap and all higher-order kinematical $D$-vectors are defined for the motion of a massive particle in a $D$-dimensional Minkowski spacetime. We illustrate the formalism with  stationary motions,  for which we provide a new, Lorentz covariant,  classification. We generalize some cases to branes, explaining the relevance to uniform motion in a heat bath.  We also consider some non-stationary motions, including motion with constant proper jerk, and free fall into a black hole as viewed from a GEMS perspective. 
}


\begin{document}

\section{Introduction}
\setcounter{equation}{0}

In the non-relativistic mechanics of a particle moving in a Euclidean space with cartesian coordinates ${\bf x}$, the particle's position at time $t$ is ${\bf x}(t)$, and  its velocity, acceleration, jerk and snap are defined by 
\be\label{vetc}
{\bf v} = \frac{d{\bf x}}{dt}\, , \qquad {\bf a} = \frac{d {\bf v}}{dt}\, , 
\qquad {\bf j} = \frac{d{\bf a}}{dt}\, , \qquad {\bf s} = \frac{d{\bf j}}{dt}\, . 
\ee
The names `crackle' and `pop' have been suggested for the next in the series. Since the discovery that the expansion of the universe is accelerating, there has been considerable interest  in  `cosmic'  jerk and snap (see e.g. \cite{Visser:2003vq,Dunajski:2008tg}).  On a more mundane level, jerk and snap are relevant in various physical or engineering contexts involving time-dependent acceleration, such as the mechanical shocks due to earthquakes or the design of  roller-coasters.  Although these might seem rather specialized areas, it should be appreciated that  jerk, snap, and all higher time derivatives are non-zero in many of  the simplest  physical contexts.  Consider the example of  an object dropped from rest in the earth's  gravitational field. Prior to being dropped it has  zero acceleration (with respect to an earth-fixed frame)  but soon afterwards has an acceleration of magnitude $g$, implying a non-zero jerk at some intermediate time; but its jerk was originally zero too, so the snap was also non-zero at some intermediate time, although it too was initially zero.  A continuation of this argument {\it ad infinitum} shows that {\it all} time derivatives of the acceleration are relevant to the simple process of releasing an object from rest. 

In view of these observations, it seems remarkable that the {\it relativistic} generalization of jerk, snap, etc. has attracted almost no attention in more than a century since the foundation of special relativity.  It might be supposed that this is because there is little new to relativistic kinematics once one has defined the  $D$-acceleration $A$, in a $D$-dimensional Minkowski spacetime, as the proper-time derivative of the $D$-velocity $U$:
\be
A = \frac{dU}{d\tau}  = \gamma \, \frac{dU}{dt}\, , \qquad \gamma= 1/\sqrt{1-v^2}\, . 
\ee
In particular, it is natural to suppose that one should define the relativistic jerk as $J= dA/d\tau$. However,  $J$ is not necessarily spacelike.  This was pointed out in our previous paper  \cite{Russo:2008gb}  and it led us to define  the relativistic jerk as   
\be
\Sigma = J  -A^2 U \ ,\qquad  J = dA/d\tau\, . 
\ee
Observe that  $U\cdot   \Sigma \equiv 0$, which implies that $\Sigma$ is spacelike if non-zero. 

Following the posting in the archives of the original version of this paper, it was brought to our attention that 
 relativistic jerk arises naturally in the context of the Lorentz-Dirac equation, which describes the motion of a charged particle in an external force field after account is taken of the radiation-reaction force due to energy loss by radiation  (see e.g. \cite{Poisson:1999tv}). In our notation, the radiation-reaction $4$-force takes the very simple form
\be
f_{\rm rad} =  - {2e^2\over 3} \Sigma  \, . 
\ee
In higher dimensional spacetimes, higher order derivatives of the acceleration play a role in  the radiation reaction problem  \cite{Galtsov:2001iv}. Radiation reaction due to gravitational wave radiation  is currently an active topic because of its importance to gravitational wave detection; see e.g. \cite{Rosenthal:2006iy}.

The main aim of this paper is to complete the formulation of relativistic kinematics by proposing adequate definitions for relativistic snap and all higher-order kinematical $D$-vectors. 
Firstly, we should address the issue of what is meant by ``proper snap''.  One natural definition  of proper acceleration, jerk and snap is
\be
{\bf a}_{prop} := \frac{d}{d\tau} {\bf v}\, \qquad {\bf j}_{prop} := \frac{d^2}{d\tau^2} {\bf v} \, , \qquad
{\bf s}_{prop} := \frac{d^3}{d\tau^3} {\bf v} \, . 
\ee
A nice feature of this definition for acceleration and jerk is that 
\be
{\bf a}_{prop} = \left. {\bf a}\right|_{v=0}\, \qquad {\bf j}_{prop} = \left. {\bf j}\right|_{v=0}\, , 
\ee
where ${\bf a}$ and ${\bf j}$ are defined as {\it coordinate-time} derivatives of velocity, as in (\ref{vetc}).  In other words, proper acceleration and proper jerk are just the particle's acceleration and jerk in the instantaneous rest-frame. 
In contrast, 
\be\label{props}
{\bf s}_{prop} =  \left( {\bf s} + a^2 {\bf a}\right)\!|_{v=0} \, , 
\ee
where ${\bf s}$ is defined as in (\ref{vetc}); the $a^2{\bf a}$  term appears because
\be\label{div2gam}
\left(d^2\gamma/dt^2\right)\!|_{v=0} = \left(a^2\right)\!|_{v=0} \, . 
\ee
As $dt=d\tau$ in the rest-frame,  we could instead define the proper snap to be $\left. {\bf s}\right|_{v=0}$.  Thus, interpretations of the term `proper'  that yield equivalent definitions for proper acceleration and proper  jerk yield  inequivalent definitions for proper snap.

A similar difficulty arises in the definition of relativistic $D$-snap.  To begin with, there is the familiar problem that the naturally defined $D$-vector $S= d\Sigma/d\tau$ is not necessarily spacelike,  but this is easily remedied by defining the relativistic snap to be the $D$-vector 
\be
\Xi=S - (A\cdot\Sigma)U \, , \qquad S= d\Sigma/d\tau\, . 
\ee
Observe that $U\cdot\Xi\equiv0$, which implies that $\Xi $ is spacelike if non-zero. However, whereas
\be
\left. A\right|_{v=0} = \left. {\bf a}\right|_{v=0} \cdot \partial_{\bf x}\, , \qquad 
\left. \Sigma\right|_{v=0} = \left. {\bf j}\right|_{v=0} \cdot \partial_{\bf x}\, , 
\ee
one finds that 
\be\label{xivzero}
\left. \Xi\right|_{v=0} = \left({\bf s} + 3a^2 {\bf a} \right)\!|_{v=0}\cdot \partial_{\bf x}\, . 
\ee
Notice that the $a^2{\bf a}$ term does {\it not} combine with ${\bf s}$ to give ${\bf s}_{prop}$ of (\ref{props}). 
One might be tempted to remedy this apparent deficiency by adding a term proportional to $A^2 A$ to the definition 
of relativistic snap because the new candidate $D$-vector for relativistic snap would still be orthogonal to $U$. 
However, this would  violate the  basic requirement that relativistic $D$-snap should vanish on  worldlines for which 
$\Sigma\equiv 0$.  This  fact  suggests  that we should define the proper snap as $\left({\bf s} + 3a^2 {\bf a} \right)\!|_{v=0}$, but we leave this point open for future debate because our results will not depend on the choice of what to call proper snap.  Similar issues arise for crackle and pop, but snap illustrates well the generic case and the extension to  crackle, pop and beyond is  straightforward. 

When considering the extension to all orders it is convenient to  denote by $P_n$ the kinematical quantity involving the $n$th proper-time derivative of the $D$-velocity, so that   $P_1=A$, $P_2=\Sigma$ and $P_3=\Xi$.  It turns out that there is a unique extension 
that satisfies the two requirements 
\be
(i) \qquad P_n^2 \ge0 \qquad (n\ge1)\, , \qquad \qquad (ii) \qquad P_n\equiv0 \Rightarrow P_{n+1}\equiv 0\, . 
\ee
The `naive' definition $P_n =d^nU/d\tau^n$ satisfies (ii) but not (i), and if one remedies this by using the projection operator 
 $(1 + UU)$ to project out the component parallel to $U$  then (ii) is satisfied, by construction,  but  not (i) when $n\ge3$. Assuming that
no dimensionful constants are to be introduced, the definition that we propose here is the unique possibility (up to scale) consistent with requirements (i) and (ii) for all $n\ge1$. 
 
 We illustrate the formalism with various examples.   In particular, we consider motions in $D$-dimensional Minkowski spacetime that are {\it stationary} in the sense that the particle's worldline is the orbit of some timelike Killing vector field. A much-studied case is that of constant proper acceleration  because of its association with the Unruh temperature $T_U=(a\hbar)/(2\pi c)$  \cite{Unruh:1976db}. The stationary motions for $D=4$ were classified in  \cite{Letaw:1980yv} using a Frenet-Serret framework in which one requires the extrinsic curvature, torsion and `hyper-torsion'  of the worldline to be constant. This classification leads to essentially six distinct possibilities (see \cite{Rosu:2000ki,Louko:2006zv} for summaries of this result, which has been extended to $D=5,6$ in \cite{Iyer}).  Using our formalism we are able to present a Lorentz covariant version of this classification. As we shall show,  it is a general 
feature of stationary motions that all $P_n$ are linear combinations of  $\{P_1,\dots,P_{D-1}\}$. For $D=4$ this means that only the acceleration, jerk and snap can be linearly independent (for stationary motions) and these quantities constitute relativistic generalizations of the curvature, torsion and hyper-torsion of the Frenet-Serret approach.  After the original version of this paper appeared on the archives, it was brought to our attention that our classification parallels, for $D=4$, a 1948 classification by Taub of the motions of a particle in a constant electromagnetic field \cite{Taub:1948zza}. This connects the kinematics of stationary motion to the dynamics of  particles in electromagnetic fields. 

Another aim of this paper is to extend the relativistic kinematics of particles to the relativistic kinematics of branes, as was done for velocity, acceleration and jerk in  \cite{Russo:2008gb}. We used there the fact that the brane's motion defines,  for a  given Lorentz frame, a congruence of timelike worldlines, for each of which one may compute (in principle) the acceleration and jerk, and now too the snap and all higher time derivatives. We shall say that the brane's motion is stationary if  all the worldlines of the congruence represent stationary particle motions. This brane perspective unifies some motions that are considered distinct  in the classification of \cite{Letaw:1980yv}  because the type of stationary motion may depend on the Lorentz frame. 

An example is a brane undergoing constant uniform acceleration; in a boosted Lorentz  frame each of the worldlines of the congruence represents a particle on the accelerating brane that is also  `drifting'  at some non-zero constant velocity ${\bf v}$ orthogonal  to the direction of acceleration.  This example provides a new way to study the effect on a particle detector of constant velocity motion in a heat bath,  by relating the  brane's  acceleration to  its temperature via the Unruh formula, although we here consider only the classical kinematics aspects of this problem.  
The fact that the jerk is non-zero for acceleration with non-zero drift velocity  implies a deviation from thermality when $v\ne0$. The deviation may be measured by the dimensionless parameters
\be\label{parameters}
\lambda = |\Sigma|/A^2\, , \qquad \eta = |\Xi|/|A|^3\, , 
\ee
the first of which was introduced in  our earlier paper \cite{Russo:2008gb}, where we argued that a necessary condition for the approximate validity of the Unruh formula is $\lambda\ll1$. This was also sufficient for the examples considered in \cite{Russo:2008gb}, and this criterion has been used since in \cite{Paredes:2008cr}.  In  general, we expect the necessary and sufficient condition to be $\lambda^{(n)}\ll1$ for all $n>0$, where 
\be\label{params}
\lambda^{(n)} = |P_{n+1}|/ |A|^{n+1}\, . 
\ee
Note that $\lambda^{(1)}=\lambda$ and $\lambda^{(2)}=\eta$. For the case of acceleration with drift,  we find that $\lambda^{(n)}=v^n$, so $\lambda\ll1$ implies $\lambda^{(n)}\ll1$ for all $n$.  

We also consider some instructive examples of non-stationary particle motion. These include a particle with constant proper jerk,  and another example in which the jerk, snap and all higher-derivatives of acceleration become progressively more important, relative to acceleration,  as a particle approaches a state of rest.   

Finally,  we consider non-stationary motions that arise in the GEMS approach to black hole thermodynamics \cite{Deser:1998bb}. In this approach, the local temperature of a stationary observer is interpreted  as an Unruh temperature due to acceleration in a higher-dimensional {\it flat} spacetime in which the black hole spacetime is globally embeded.  Here we consider test particles that are freely falling in the black hole metric, as in 
\cite{Brynjolfsson:2008uc} but we address some additional  issues.  In particular, we consider what happens 
as a particle  approaches the singularity of  a Schwarzschild black hole, after it has fallen through the event 
horizon.  Apart from providing insights into the nature of black hole singularities, this example also provides a natural class of examples in which the acceleration and jerk increase without bound.


\section{Jerk, Snap, and all that}
\label{sec:kinematics}
\setcounter{equation}{0}

We first recall the essential features of the relativistic mechanics of a particle in a Minkowski spacetime of dimension $D$. The Minkowski metric is  
\be\label{Minkmetric}
ds^2 = dX^\mu dX^\nu \eta_{\mu\nu}  = -dt^2 + d{\bf x} \cdot d{\bf x}\, . 
\ee
The particle's worldline is specified by the functions $X^\mu(t)$ and its proper-time differential is 
\be
d\tau = \sqrt{-ds^2} = \gamma^{-1} dt\, , \qquad \gamma = \frac{1}{\sqrt{1-v^2}}\, . 
\ee
The particle's $D$-velocity and $D$-acceleration are, respectively, 
\begin{eqnarray} 
U &\equiv&  U^\mu \partial_\mu \, , \qquad 
U^\mu = \dot X^\mu= \gamma  dX^\mu /dt \nonumber\\
A &\equiv& A^\mu \partial_\mu\, , \qquad A^\mu = \dot U^\mu = \gamma  dU^\mu /dt ,
\end{eqnarray}
where the overdot indicates differentiation with respect to proper time $\tau$. 
A calculation yields
\be\label{Acalc}
U =  \gamma \partial_t + \gamma {\bf v}\cdot \partial_{\bf x}\, , \qquad 
A= \gamma^4 \left({\bf v}\cdot {\bf a} \right) \partial_t + \gamma^2\left[{\bf a} + 
\gamma^2\left({\bf v}\cdot{\bf a}\right) {\bf v} \right]\cdot\partial_{\bf x}\, . 
\ee
Observe that
\be
U^2=-1\, , \qquad U\cdot A =0\, , \qquad A^2 = \gamma^4\left[a^2 + \gamma^2({\bf v}\cdot {\bf a})^2\right] = a^2|_{v=0}\, . 
\ee
The acceleration ${\bf a}$ appearing in this formula is the $(D-1)$-acceleration of the particle, at a given time $t$,  in the frame in which its $(D-1)$-velocity at that time is ${\bf v}$. In particular $A= {\bf a} \cdot \partial_{\bf x}$ in the instantantaneous rest frame, defined as choice of frame for which  ${\bf v}={\bf 0}$. This is standard, see e.g. \cite{Rindler}, but it is important to appreciate that ${\bf a}$ in the instantantaneous rest frame is not the same as ${\bf a}$ in the `laboratory'  frame. To stress this fact, we prefer to write ${\bf a}$ in the instantantaneous rest frame as ${\bf a}|_{v=0}$. Thus, 
\be
U|_{v=0} = \partial_t \, , \qquad  A|_{v=0} = \left. {\bf a}\right|_{v=0}\cdot\partial_{\bf x}\, .  
\ee
Performing a Lorentz transformation from the instantaneous rest-frame to the frame in which the particle has (instantaneously) velocity ${\bf v}$, we recover
$A$ of (\ref{Acalc}) with
\be
{\bf a} = \gamma^{-3} \left[\gamma\,  {\bf a}|_{v=0} + \frac{\left(1-\gamma\right)}{v^2}  \left({\bf v}\cdot {\bf a}|_{v=0}\right) {\bf v}\right] \, , 
\ee
which is the standard  Lorentz transformation of the 3-acceleration from the instantaneous rest-frame to the `laboratory' frame.

\subsection{Relativistic Jerk}

It would be natural to try to define the relativistic jerk as (e.g. \cite{MM})
\begin{eqnarray}
J = \frac{d A}{ d\tau} &=&  \gamma^5\left[a^2 + {\bf v}\cdot{\bf j}+ 4\gamma^2 ({\bf v}\cdot{\bf a})^2\right] \partial_t 
\nonumber \\
&& + \  \gamma^3 \left\{ {\bf j} + 3\gamma^2({\bf v}\cdot {\bf a}) {\bf a}  +  \gamma^2\left[a^2 + {\bf v}\cdot{\bf j} + 4\gamma^2({\bf v}\cdot{\bf a})^2\right]{\bf v} \right\}\cdot\partial_{\bf x}\, . 
\end{eqnarray}
However, in the instantaneous rest frame this becomes
\be
J_{v=0} = \left(a^2\right)|_{v=0} \partial_t + \left. {\bf j}\right|_{v=0} \cdot \partial_{\bf x}\, , 
\ee
so $J$ is not  necessarily spacelike, and may even be non-zero when $j_{v=0}=0$.   It makes more sense to define the relativistic jerk to be \cite{Russo:2008gb}
\be
\Sigma = J -A^2 U \, , 
\ee
because this satisfies the identity
\be
U\cdot \Sigma \equiv 0\, , 
\ee
which  implies that $\Sigma$ is  spacelike. A computation shows that
\begin{eqnarray}
\Sigma &=&  \gamma^5\left[ {\bf v}\cdot{\bf j}+ 3\gamma^2 ({\bf v}\cdot{\bf a})^2\right] \partial_t 
\nonumber \\
&& + \  \gamma^3 \left\{ {\bf j} + 3\gamma^2({\bf v}\cdot {\bf a}) {\bf a}  +  \gamma^2\left[ {\bf v}\cdot{\bf j} + 3\gamma^2({\bf v}\cdot{\bf a})^2\right]{\bf v} \right\}\cdot\partial_{\bf x}\, , 
\end{eqnarray}
from which we see that
\be
\Sigma_{v=0} =  \left. {\bf j}\right|_{v=0} \cdot \partial_{\bf x}\, .  
\ee
Furthermore, 
\be\label{Sigtoj}
\Sigma =0 \quad \Leftrightarrow \quad {\bf j} = -3\gamma^2 \left({\bf v}\cdot{\bf a}\right) {\bf a}\, , 
\ee
so that $\Sigma=0$ in the instantaneous rest-frame whenever ${\bf j}={\bf 0}$ in this frame, and {\it vice-versa}. We note here, for future use, that 
\be\label{jtos}
{\bf j} = -3\gamma^2 \left({\bf v}\cdot{\bf a}\right) {\bf a}\ \quad \Rightarrow \quad
{\bf s} = -3\gamma^2\left[a^2 -4\gamma^2\left({\bf v}\cdot{\bf a}\right)^2\right] {\bf a}\, . 
\ee

\subsection{Relativistic Snap}

We now come to the relativistic generalization of snap. One possibility, would be to define it as
\begin{eqnarray}
S = \frac{d\Sigma}{d\tau} &=& \gamma^6 \left\{ {\bf v}\cdot{\bf s} + {\bf a}\cdot{\bf j} + \gamma^2 ({\bf v}\cdot{\bf a}) \left(6a^2 + 11{\bf v}\cdot{\bf j}\right) + 21 \gamma^4 ({\bf v}\cdot{\bf a})^3 \right\}\partial_t \nonumber \\
&+& \gamma^4\left\{ {\bf s} + 6\gamma^2 ({\bf v}\cdot{\bf a}){\bf j} + \gamma^2\left[3a^2+4{\bf v}\cdot{\bf j} +
18\gamma^2({\bf v}\cdot{\bf a})^2\right] {\bf a} \right.  \\
&& \left. + \gamma^2\left[{\bf v}\cdot{\bf s} + {\bf a}\cdot{\bf j} + \gamma^2 ({\bf v}\cdot{\bf a}) \left(6a^2 + 11{\bf v}\cdot{\bf j}\right) + 21 \gamma^4 ({\bf v}\cdot{\bf a})^3 \right] {\bf v}\right\}\cdot \partial_{\bf x}\ .
\nonumber
\end{eqnarray}
However, this suffers from the same problems as  $J$. In particular, 
\be
S|_{v=0} = ({\bf a}\cdot {\bf j})|_{v=0} \, \partial_t + \left( {\bf s} + 3 a^2{\bf a}\right)\!|_{v=0} \cdot\partial_{\bf x}\, , 
\ee
so $S$ is not  necessarily spacelike. This problem is easily resolved following the jerk example: we instead define the particle's relativistic snap to be
\be
\Xi = S - \left(A\cdot\Sigma\right) U \equiv  \frac{d\Sigma}{d\tau} - \left(A\cdot\Sigma\right) U\, . 
\ee
This satisfies the identity
\be
U\cdot \Xi \equiv 0\, , 
\ee
which implies that $\Xi$ is spacelike. A calculation shows that
\begin{eqnarray}
\Xi &=& \gamma^6 \left\{ {\bf v}\cdot{\bf s}  + \gamma^2 ({\bf v}\cdot{\bf a}) \left(3a^2 + 10{\bf v}\cdot{\bf j}\right) + 18\gamma^4 ({\bf v}\cdot{\bf a})^3 \right\}\partial_t \nonumber \\
&+& \gamma^4\left\{ {\bf s} + 6\gamma^2 ({\bf v}\cdot{\bf a}){\bf j} + \gamma^2\left[3a^2+4{\bf v}\cdot{\bf j} +
18\gamma^2({\bf v}\cdot{\bf a})^2\right] {\bf a} \right.  \\
&& \left. + \gamma^2\left[{\bf v}\cdot{\bf s}  + \gamma^2 ({\bf v}\cdot{\bf a}) \left(3a^2 + 10{\bf v}\cdot{\bf j}\right) + 18 \gamma^4 ({\bf v}\cdot{\bf a})^3 \right] {\bf v}\right\}\cdot \partial_{\bf x}\, , \nonumber
\end{eqnarray}
so that
\be
\Xi|_{v=0} = \left( {\bf s} + 3a^2{\bf a}\right)\!|_{v=0} \cdot \partial_{\bf x}\, . 
\label{uuu}
\ee
We thus confirm that $\Xi$ is spacelike, although it has the curious feature that it may be non-zero in the rest-frame even when ${\bf s}$ vanishes in this frame. 

As pointed out in the introduction, there is an ambiguity in what one might mean by ``proper snap'', so the fact that $\Xi|_{v=0} \ne {\bf s}|_{v=0} \cdot \partial_{\bf x}$ is not really a problem.  In any case, this feature is unavoidable; suppose that we were to try to  `improve'  upon our definition of relativistic snap by  considering
\be\label{hatXi}
\Xi_\alpha = \Xi - \alpha \ A^2 A\, ,
\ee
for some constant $\alpha$. We now have a one-parameter family of  $D$-vectors orthogonal to $U$, each of which could be considered a candidate for relativistic $D$-snap. However, $\Xi_\alpha$ {\it does not vanish identically for a particle with $\Sigma \equiv 0$ } (for which $\Xi\equiv 0 $) unless $\alpha=0$, and this makes any other choice 
unacceptable, in our view.

\subsection{Crackle, Pop and beyond: $P_n$} 

To go on to higher derivatives of the acceleration we need an appropriate notation. We 
will use the notation 
\be
P_0 \equiv U\, , \qquad P_1 \equiv A\, \qquad  P_2 \equiv \Sigma \, , \qquad P_3 \equiv \Xi\, , 
\ee
and we define all higher time derivatives in terms of the lower ones via the formula
\be\label{gar}
P_{n+1}= \frac{dP_n}{d\tau} - (A\cdot P_n)\ U\, . 
\ee
Observe that this formula is consistent with our previous definitions of relativistic jerk and snap, and it extends these definitions to all higher orders in time derivatives. From this definition it follows that 
\be
U\cdot P_n \equiv 0  \qquad \left(n>0\right), 
\label{jujuy}
\ee
and hence that all non-zero $P_n$ with $n>0$ are spacelike. In addition, 
\be
P_n \equiv 0 \quad \Rightarrow \quad P_{n+1}\equiv 0\, , 
\label{salta}
\ee
irrespective of the values of $\left\{P_1,\dots,P_{n-1}\right\}$.  

The properties (\ref{jujuy}) and (\ref{salta}) fix the definition uniquely once one specifies that 
\be
P_n\ \big|_{v=0}  =  \left[\left(\frac{d^n}{dt^n} {\bf v}\right) + \dots\right] _{v=0} \cdot \partial_{\bf x} \, , 
\ee
where the ellipsis indicates a possible sum (present for $n>2$) of terms involving only lower derivatives of ${\bf v}$. 
This can also be written as 
\be
P_n\ \big|_{v=0}  =  \left[\left(\frac{d^n}{d\tau^n} {\bf v}\right) + \dots\right]_{v=0}  \cdot \partial_{\bf x} \, , 
\ee
with a different sum over lower derivative terms. Here is a list of the first few $P_n$:
\begin{eqnarray}
 {\rm velocity:}  & P_0 & \equiv  U \nonumber \\
 {\rm acceleration:}  & P_1 &\equiv A = \frac{dU}{d\tau} \nonumber\\
 {\rm jerk:}  &  P_2 &\equiv \Sigma = \frac{dA}{d\tau} - A^2\,  U  \nonumber\\
 {\rm snap:} & P_3 &\equiv \Xi = \frac{d\Sigma}{d\tau} - \left(A\cdot\Sigma\right) U  \nonumber\\
 {\rm crackle:}  & P_4 &\equiv C = \frac{d\Xi}{d\tau} - \left(A\cdot \Xi\right) U
\end{eqnarray}

Just as we considered the family of $D$-vectors $\Xi_\alpha$ of (\ref{hatXi}) as possible alternative definitions of snap, so we may also consider $P_n$ for $n>3$ as a special case of a  familiy of $D$-vectors. For example
one may define a $D$-vector  $\hat P_n$ with the property that 
\be\label{pens}
\hat P_n\ \big|_{v=0} = \left(\frac{d^n}{dt^n} {\bf v}\right) \cdot \partial_{\bf x} \, .
\ee
One has $\hat P_n= P_n$ for $n=0,1,2$, but starting with  $n=3$ (snap) we would find a deviation from our 
preferred definition (\ref{gar}). Specifically, one can show that the $\hat P_n$  are defined, iteratively, by the formula
\be
\hat P_{n+1} = \frac{d \hat P_n}{d\tau} - (A\cdot \hat P_n)\, U 
- \sum_{k=1}^{n-1}   c_{n,k} (A\cdot \hat P_{k})\,  \hat P_{n-k}
\label{agaf}
\ee
where
\be
c_{n,k}= {(n+1)!\over (n-k)! (k+1)!}\, . 
\ee 
However, for $n\ge3$ the $D$-vectors $\hat P_n$ do not have the property  that 
$P_{n+1} \equiv0$ when $P_n\equiv 0$. The same 
goes for all variants of $\hat P_n$. 

\section{Stationary Motions}
\setcounter{equation}{0}

The motion of a particle is stationary if 
\be
U =k
\ee
for a Killing vector field $k$ normalized, locally on the particle's worldline, such that $k^2=-1$ (for a particle with non-zero rest-mass).  For $D$-dimensional Minkowski space with metric (\ref{Minkmetric}), all Killing vector fields have the form
\be
k = b_0\, \partial_t + {\bf b}\cdot \partial_{\bf x} + {\bf \tilde a} \cdot \left(t\partial_{\bf x} + {\bf x}\partial_t\right) + 
\left(\Omega {\bf x}\right) \cdot \partial_{\bf x}\, , 
\ee
where $b_0 $ is a constant, ${\bf b}$ and ${\bf \tilde a}$ are constant $(D-1)$-vectors,  and $\Omega$ is a constant 
$(D-1)\times (D-1)$ antisymmetric matrix.  The condition $k^2=-1$ is equivalent to the constraint 
\begin{eqnarray}\label{KVFcon}
1 &=& b_0^2 - b^2 - 2 \left({\bf b}\cdot {\bf \tilde a}\right) t + 2\left(b_0 {\bf \tilde a} + \Omega {\bf b}\right)\cdot {\bf x} 
- \tilde a^2 \nonumber\\ 
&& -\ \tilde a^2 t^2 + 2\left(\Omega \tilde a\right) \cdot {\bf x} t - \left|\Omega {\bf x}\right|^2 + \left({\bf \tilde a}\cdot {\bf x}\right)^2\, . 
\end{eqnarray}
The equation $U=k$ is equivalent to the equations
\be\label{stationeq}
\dot t = b_0 + {\bf \tilde a} \cdot {\bf x}\, , \qquad \dot {\bf x} = {\bf b} + {\bf \tilde a} t + \Omega {\bf x}\, , 
\ee
where the overdot indicates differentiation with respect to the particle's proper time $\tau$; one may verify that the constraint (\ref{KVFcon}) is consistent with this evolution.  We may write these equations as
\be
\dot X^\mu = \xi^\mu + S^\mu{}_\nu X^\nu\, , 
\label{avaf}
\ee
where the components of the vector $\xi$ and tensor $S$ are
\be
\xi = \left( \begin{array}{c} b_0 \\  {\bf b}  \end{array} \right) \, , \qquad
S= \left( \begin{array}{cc} 0 & {\bf \tilde a} \\ {\bf \tilde a} & \Omega \end{array} \right) \, . 
\ee
When $S=0$ the motion is inertial. When  $S$ is non-zero and diagonalizable, the  type of stationary motion depends on the qualitative properties of its eigenvalues and, in the case of zero eigenvalues, the components of $\xi$ projected onto the kernel of $S$.  The cases for which $S$ is non-zero but non-diagonalizable must be treated separarately.

For stationary motions all Lorentz invariants constructed from the $D$-velocity and its derivatives must be constant. In particular,  $P_n^2$ is constant for all $n$ and this implies that $P_n\cdot P_{n+1} =0$. It can also be proved by induction that $dP_n/d\tau = S P_n$ for $n\ge0$, and hence that  
\be\label{Sform}
P_{n+1} = SP_n -\left(A\cdot P_n\right) U\, . 
\ee
One can also show that 
\be\label{recursion2}
A\cdot P_{2m}=0 \, , \qquad A\cdot P_{2m+1} = \left(-1\right)^{m} P_{m+1}^2\, . 
\ee
This result, together with (\ref{Sform}) implies the recursion relations
\be\label{recursion}
P_{2m+1}= SP_{2m}\ , \qquad P_{2m+2}= S P_{2m+1} - \left(-1\right)^m P_{m+1}^2\,  U\, , \qquad (m\ge0).
\ee
We will sketch the proof of (\ref{recursion2}).   Using (\ref{Sform}) for $n=2m$,  we have 
\be
A\cdot P_{2m} = A \cdot\left[SP_{2m-1}\right] = - \left(SA\right) \cdot P_{2m-1}\, , 
\ee
where we use the orthogonality of $A$ with $P_{2m}$ for the first equality and the antisymmetry of $\eta S$ for the second one. Now using (\ref{Sform}) for $n=2$ in the form $SA= P_2 + A^2U$, we deduce that
\be
A\cdot P_{2m}  = -P_2 \cdot P_{2m-1}\, . 
\ee
We can now repeat the process by using (\ref{Sform}) for $n=2m-1$ to show that  $P_2 \cdot P_{2m-1} =- P_3 \cdot P_{2m-2}$. Further repetition of the same argument leads to the conclusion that
\be
A\cdot P_{2m}  = -P_2 \cdot P_{2m-1} = P_3 \cdot P_{2m-2} = \cdots = P_m \cdot P_{m+1} =0\, . 
\ee
A similar argument shows that
\be
A\cdot P_{2m+1} = - P_2 \cdot P_{2m} = P_3 \cdot P_{2m-1} = \cdots = \left(-1\right)^m P_{m+1}^2 \, . 
\ee
The recursion relations (\ref{recursion}) have the solution 
\be
S^n U = P_n +  \sum_{k=1}^{[n/2]} \left(-1\right)^{k+1} P_k^2 \, S^{n-2k}U
\label{snn}
\ee
where $[n/2]$ is the integer part of $n/2$. It follows that   $S^nU$ ($n>0$) is a linear combination of $(P_1,\dots P_n)$ with constant coefficients. Conversely, $P_n$ ($n>0$) is a linear combination of $(SU,\dots, S^nU)$ with constant coefficients, the coefficient of the  $S^nU$ term being unity.  However, the Cayley-Hamilton theorem implies that $S$ satisfies a polynomial  constraint of order $D$, and hence that the elements of the set $(U,SU,\dots, S^nU)$ must be linearly 
dependent  when $n\ge D$. Because $U$ is timelike and  all other non-zero vectors of this set are spacelike,  the same statement is  true of the smaller set $(SU,\dots, S^nU)$.   It follows that $P_n$ for $n\ge D$ is a linear combination of  the $P_k$ with $k<D$, with constant coefficients\footnote{Of course there are only $D$ independent directions,  but in the general case of  non-stationary motion  all $P_n$ are independent in the sense than none can be expressed, generically,  as a linear combination of others with constant coefficients. }.

We shall analyse in detail the $D=4$ case, verifying that all $P_n$ are linear combinations of $(A,\Sigma,\Xi)$. 
In this case we may write
\be
\Omega_{ij} =  \varepsilon_{ijk} \omega_k\, , 
\ee
where $\omega_k$ are the components of a $3$-vector  {\boldmath $\omega$}, with length $\omega$.  It is convenient to define the Lorentz invariants
\be\label{LI}
I_1 := \frac{1}{2}{\rm tr} S^2 = \tilde a^2-\omega^2\, , \qquad 
I_2 :=  \frac{1}{4}{\rm tr} S^4 - \frac{1}{8} \left({\rm tr} S^2\right)^2 = \left({\bf \tilde a} \cdot  \mbox{\bfom}\right)^2 \, . 
\ee
If both these invariants vanish for non-zero $S$ then $S$ is not diagonalizable. Otherwise, there are four eigenvectors with eigenvalues $e$ given by
\be\label{lamform}
2e^2 =  I_1 \pm \sqrt{I_1^2 + 4I_2}\, .
\ee
In the following subsections we shall go through the various possibilities. 

As mentioned in the introduction, this method of classifying stationary motions is the same as that used by 
Taub to classify the motions of a charged particle in a constant electromagnetic field \cite{Taub:1948zza}.
For a particle of electric charge $e$ and mass $m$,  the Lorentz force law yields the equation of motion
\be
\ddot X^\mu = {e\over m} \ F^\mu {}_\nu \dot X^\nu \ ,
\ee
where $F_{\mu\nu} = \eta_{\mu\rho}F^\rho{}_\nu$ are the components of  the electromagnetic $2$-form $F$. 
When $F$ is constant this equation can be integrated once to get our equation (\ref{avaf}) with
\be
S^\mu{}_\nu = (e/m) F^\mu{}_\nu\, . 
\ee
It follows that  $I_1$ and $I_2$ are, for $D=4$,  proportional to the familiar Lorentz invariants  $E^2-B^2$ 
and $(E\cdot B)^2$, respectively. Naturally, there are more such invariants in higher dimensions. 
For any dimension $D$, this shows (i) that each stationary motion arises as the  trajectory of a charged particle in a particular constant electromagnetic background, and (ii) all motions of a charged particle in a constant electromagnetic background are stationary. 

\subsection{Acceleration plus rotation; generic case}

First we consider $I_2 \ne0$. In this case $S$ is invertible so we may  choose $\xi=0$ without loss of generality. 
Two of the eigenvalues of $S$ are real and two are imaginary. We may choose  a Lorentz frame in 
which $\tilde {\bf a}$ and $\mbox{\bfom}$ are parallel, so that  $\tilde{\bf a} \times \mbox{\bfom} ={\bf 0}$; for example
$\tilde {\bf a} =(\tilde a,0,0)$ and  $\mbox{\bfom} = (\omega,0,0)$.  The four eigenvalues of $S$ are then
$(\tilde a, -\tilde a,  i\omega, -i\omega)$. Using the notation ${\bf x}= (x,y,z)$,  one finds that
\begin{eqnarray}\label{traj}
t &=& \tilde a^{-1} \sqrt{1+R^2\omega^2}\, \sinh\left(\tilde a\tau\right)\, , \qquad 
x= \tilde a^{-1} \sqrt{1+R^2\omega^2}\, \cosh\left(\tilde a\tau\right) \nonumber \\
y &=& - R \cos\left(\omega\tau\right) \, , \qquad
z  =  R \sin\left(\omega\tau\right)\, , 
\end{eqnarray}
where $R $ is an arbitrary constant.  The particle moves in a planar circle  with constant proper angular frequency $\omega$, and it accelerates  along the normal to the plane  with constant proper acceleration 
\be
a = \tilde a \sqrt{1+ R^2\omega^2}\, . 
\ee
The $4$-velocity and $4$-acceleration are 
\begin{eqnarray}
U &=& \sqrt{1+R^2\omega^2} \left[\cosh \left(\tilde a\tau\right) \partial_t + \sinh \left(\tilde a\tau\right)\partial_x \right]
+ R\, \omega \left[\sin\left(\omega\tau\right) \partial_y + \cos\left(\omega\tau\right) \partial_z\right] \nonumber\\
A &=& a \left[\sinh\left(\tilde a\tau\right)\partial_t + \cosh\left(\tilde a\tau\right)\partial_x\right] 
+ R\, \omega^2 \left[\cos\left(\omega\tau\right)\partial_y  - \sin\left(\omega\tau\right)\partial_z \right] \, . 
\end{eqnarray}
Note that
\be
A^2 = a^2 + \left(R\omega^2\right)^2 \, . 
\ee
A computation of the jerk yields 
\be
\Sigma = - |\Sigma|\left\{ R\, \omega \left[\cosh \left(\tilde a\tau\right) \partial_t + \sinh \left(\tilde a\tau\right)\partial_x \right] + 
\sqrt{1+R^2\omega^2} \left[\sin\left(\omega\tau\right) \partial_y + \cos\left(\omega\tau\right) \partial_z\right]\right\}
\ee
where
\be
|\Sigma | = R\, \omega \sqrt{1+R^2\omega^2}\left(\tilde a^2+\omega^2\right)\, . 
\ee
Since $A\cdot\Sigma\equiv 0$, it is simple to compute the relativistic snap:
\be
\Xi = -|\Sigma| \omega \left\{ \tilde a R \left[\sinh\left(\tilde a\tau\right)\partial_t + \cosh\left(\tilde a\tau\right)\partial_x\right]  
+ \sqrt{1+R^2\omega^2}  \left[\cos\left(\omega\tau\right)\partial_y  - \sin\left(\omega\tau\right)\partial_z \right] \right\}
\ee
Observing that $A\cdot\Xi= -|\Sigma|^2$, we proceed to compute the relativistic crackle. The result is
\be
C= - \omega^2 \left[1 + R^2\left(\tilde a^2+\omega^2\right)\right] \Sigma\, . 
\ee
In other words, the crackle is proportional to the jerk. It follows that all higher time-derivatives are proportional to 
either the jerk or the snap, and all vanish when $R =0$. Thus, the acceleration, jerk and snap, determine the motion: the constants $(a,\omega,R)$ that define the trajectory (\ref{traj}) can be expressed in terms of  $(A^2,\Sigma^2,\Xi^2)$.

\subsection{Helical motion}

For $I_2=0$ and non-zero $I_1$ there are two cases to consider, according to the sign of $I_1$. We first consider the $I_1<0$ case, for which two of the eigenvalues of $S$ are zero and two are imaginary.   We may choose a Lorentz frame for which $\tilde a=0$, in which case the eigenvalues are $(0,0,i\omega,-i\omega)$.  The particle motion is  planar and circular, with constant proper angular velocity $\omega$. As  there are two zero eigenvalues, there are two linear combinations of the components of $\xi$ that cannot be removed by a shift of the origin, and these correspond to a constant  $2$-velocity  of the plane. Choosing $\mbox{\bfom} = (\omega,0,0)$,  and ${\bf x}= (x,y,z)$,  one finds that
\be
t=  \sqrt{1+b^2+ R^2\omega^2}\,  \tau\, , \qquad  x= b\tau\, , \qquad 
y = R\cos\left(\omega\tau\right) \, , \qquad z = R\sin\left(\omega\tau\right)\, , 
\ee
where $R$ is an arbitrary constant.  
The particle's $4$-velocity is 
\be
U = \gamma\, \partial_t + b\partial_x -
\omega R\left[\sin\left(\omega\tau\right)\partial_y - \cos \left(\omega\tau\right)\partial_z \right] \, 
\qquad \left(\gamma=  \sqrt{1+b^2 + \omega^2 R^2}\right)\, , 
\ee
and the $4$-acceleration and $4$-jerk  are
\begin{eqnarray}
A &=& - \omega^2 R \left[ \cos \left(\omega\tau\right)\partial_y + 
\sin \left(\omega\tau\right)\partial_z\right]\nonumber \\
\Sigma &=& - \omega^4 R^2 \left(\gamma\partial_t + b\partial_x\right) + 
\omega^3R\left(1+ \omega^2R^2\right)\left[ \sin \left(\omega\tau\right)\partial_y 
- \cos \left(\omega\tau\right)\partial_z\right] \, . 
\end{eqnarray}
A computation of the $4$-snap yields
\be
\Xi= -\omega^2 \left(1+ \omega^2R^2\right) A\, .
\ee
It follows that snap and all higher-derivatives are proportional to the acceleration or the jerk. In fact, 
the $P_n$ for even $n\ge2$ are given by
\be
P_{2k+2} = \left(-1\right)^k \omega^{2k} \left(1+ R^2\omega^2\right)^k \Sigma \, , \qquad k=0,1,2,\dots 
\ee
and for odd $n\ge1$ by 
\be
P_{2k+1} =  \left(-1\right)^k  \omega^{2k}\left(1+ R^2\omega^2\right)^k  A\, , \qquad k=0,1,2,\dots
\ee
Observe that these are {\it independent of the parameter $b$}; this accounts for the fact that helical motion is not distinguished from  circular motion in the classification of \cite{Letaw:1980yv}.

\subsection{Acceleration, with and without drift}
\label{subsec:Acc+}

We now consider $I_2=0$ and $I_1>0$, in which case two of the eigenvalues of $S$ are zero; the other two are real. We may choose  a Lorentz frame for which $\omega=0$, in which case the eigenvalues are $(0,0,\tilde a,-\tilde a)$. As  there are two zero eigenvalues there are two linear combinations of the components of $\xi$ that cannot be removed by a shift of the origin, and these correspond to the components of a constant `drift' velocity in the plane with normal ${\bf \tilde a}$. Without loss of generality, we may choose $b_0=0$ and 
\be
{\bf \tilde a}=\gamma (a,0,0)\, , \qquad  {\bf b}=\gamma (0,v,0) \, , \qquad \gamma =  1/\sqrt{1-v^2}\, . 
\ee
Given that ${\bf x}= (x,y,z)$, one finds that $z$ is fixed while
\be
t = a^{-1} \sinh \left(a\gamma\tau\right)\, , \qquad x= a^{-1} \cosh\left(a\gamma\tau\right)\, , \qquad 
y= \gamma v \tau\,  . 
\ee
The $4$-velocity and $4$-acceleration are 
\begin{eqnarray}
U &=& \gamma\left\{ \cosh\left(a\gamma \tau \right) \partial_t + 
\sinh\left(a\gamma \tau \right) \partial_x + v \partial_y \right\}  \nonumber \\
A &=& a\gamma^2\left\{ \sinh\left(a\gamma \tau \right) \partial_t + 
\cosh\left(a\gamma \tau \right) \partial_x \right\}\, . 
\end{eqnarray}
Observe that $|A| =  \gamma^2 a$. When $v=0$ the particle undergoes a constant proper acceleration $a$ in the $z$ direction. For non-zero 
$v$ the particle drifts with velocity $v$ in the $y$ direction and its proper acceleration in the $z$ direction becomes $\gamma^2 a$; the factor of $\gamma^2$ is a time-dilation effect. 

It is instructive to compute the jerk in two steps. First we compute
\be\label{Jcomp}
J \equiv \frac{dA}{d\tau} = a^2\gamma^3\left\{ \cosh\left(a\gamma \tau \right) \partial_t + \sinh\left(a\gamma \tau \right) \partial_x \right\}\, . 
\ee
Then we compute
\be
\Sigma \equiv J -A^2U = - v a^2\gamma^5\left\{ v \cosh\left(a\gamma \tau\right) \partial_t + v  \sinh\left(a\gamma \tau \right) \partial_x + \partial_y \right\}\, . 
\ee
Further computation reveals that
\be
\Xi = -v^2 a^2\gamma^4 A\, , \qquad C= -v^2 a^2 \gamma^4\,  \Sigma\, , 
\ee
and hence that all $P_n$ with $n\ge3$ are proportional either to $A$ or to $\Sigma$. 

Observe that $\Sigma$ is non-zero when $v\ne0$,  in spite of the fact that the proper acceleration has constant magnitude for 
any $v$. Furthermore, the non-zero component of $\Sigma$ is its $y$-component even though  $A_y\equiv0$. 
In non-relativistic mechanics, a non-zero jerk in a given direction implies that the acceleration in that direction cannot be identically zero. This would be true in relativistic mechanics too if the jerk were defined to be  $J= dA/d\tau$, and this fact might tempt one to reconsider whether $\Sigma$ really is a better definition of relativistic jerk than $J$,  but one sees from (\ref{Jcomp}) that $J^2 =- a^4 \gamma^6$ so that  $J$ is not only timelike in this example but also {\it non-zero even when $v=0$}. In contrast, $\Sigma$ vanishes identically when $v=0$ and is  at least spacelike 
when $v\ne0$.  Also, the fact that  $\Sigma \propto v$ correlates well with the fact that the  extrinsic geometry of a $v\ne0$ worldline is qualitatively different from the extrinsic geometry  of  a $v=0$ worldline, which is why these two cases are distinguished in the classification of \cite{Letaw:1980yv}. However, they are naturally considered together in our analysis, and in the following section  we shall  show that they are not really distinct from a `brane' perspective.

\subsection{Acceleration plus rotation; null case}

Finally we must consider the case in which $I_1=I_2=0$.  In this case $S^3\equiv 0$, so  $S$ is not diagonalizable. However, we may choose ${\bf \tilde a}=(a,0,0)$ and $\mbox{\bfom}=(0,0,a)$. Given that ${\bf x}=(x,y,z)$, we then find from  (\ref{stationeq}) that $z$ is constant while we may choose the origin in the $(x,y)$-plane to arrange for ${\bf b}={\bf 0}$. The solution of the resulting equations, with $b_0=1$ and the parameters subject to (\ref{KVFcon}) is  
\be
t= \tau + \frac{1}{6} a^2 \tau^3\, , \qquad x= \frac{1}{2}a \tau^2\, , \qquad y= - \frac{1}{6}a^2 \tau^3\, . 
\ee
The $4$-velocity and $4$-acceleration and $4$-jerk are
\begin{eqnarray}
U &=& \left(1+ \frac{1}{2}a^2 \tau^2\right)\partial_t + a \tau\partial_x - 
\frac{1}{2}a^2\tau^2\partial_y \nonumber\\
A&=& a^2\tau \partial_t + a\partial_x - a^2\tau \partial_y\nonumber \\
\Sigma &=& -a^2 \left[ \frac{1}{2} a^2\tau^2\partial_t + a \tau\partial_x + 
\left(1- \frac{1}{2}a^2\tau^2\right)\partial_y\right] \, . 
\end{eqnarray} 
Note that $\Sigma\cdot A=0$.
A computation of the $4$-snap yields
\be
\Xi = -a^2 A\, , 
\ee
from which it follows that all higher time derivatives are proportional either to $A$ or to $\Sigma$. 
Their explicit expressions can easily be determined  by using eq. (\ref{avaf}) iteratively and using the fact that $S^3=0$. One finds that
\be
P_{2n}=(-a^2)^{n-1}\Sigma \ ,\qquad 
P_{2n+1}= (-a^2)^n A\ .
\ee

\subsection{$D>4$ stationary motions}

The stationary motions in a  Minkowski  spacetime of dimension $D>4$ may be classified in a similar way. For even $D$ all eigenvalues of $S$ occur in pairs of opposite sign,  and this is also true for odd $D$ except for one additional zero eigenvalue.  The number of  parameters generically needed to specify the eigenvalues is therefore $[D/2]$  (the integer part of $D/2$), and these are linear combinations of the Lorentz invariants $\{ {\rm tr }S^2, {\rm tr} S^4, \dots, {\rm tr} S^{[D/2]}\}$. It is convenient to choose the linear combinations defined by the characteristic polynomial 
\be
p_S(t) \equiv \det \left(\bI\, t -S \right) =  t^D - I_1\,  t^{D-2}-I_2 \, t^{D-4} -  \ \dots\  - I_{[D/2]} \, , 
\ee
where $t$ is just a real parameter, not to be confused with time.  The Lorentz invariants $I_1$ and $I_2$ so defined coincide with those defined in  (\ref{LI}), and  the next one is
\be
I_3=  \frac{1}{6} {\rm tr} S^6 - \frac{1}{8} \left({\rm tr} S^2\right)\left({\rm tr} S^4\right) + 
\frac{1}{48}\left({\rm tr} S^2\right)^3\, . 
\ee

For $D=5$, the Lorentz invariants $I_1$ and $I_2$ suffice and are given by
\be
I_1 = a^2 + \frac{1}{2} {\rm tr} \Omega^2\, , \qquad 
I_2 = -\left[\det \Omega + \frac{1}{2} a^2 {\rm tr} \Omega^2 + \left|\Omega {\bf a}\right|^2\right] \, . 
\ee
There is always one zero eigenvalue, and the remaining four  eigenvalues are given by
\be
2e^2 = I_1 \pm \sqrt{I_1^2 + 4I_2}\, . 
\ee 
This  is formally the same as (\ref{lamform}) but now with the $D=5$ expressions for the Lorentz invariants. One important difference is that $I_2$ may now be negative, leading to the new possibility of  a rotation in two orthogonal planes. This is the only case that is essentially new, relative to $D=4$, in that it requires four space dimensions.

For $D=6$ one has
\begin{eqnarray}
I_1 &=& a^2 + \frac{1}{2} {\rm tr}\,  \Omega^2\, , \qquad 
I_2 = \frac{1}{4}{\rm tr}\,  \Omega^4 - \frac{1}{8} \left({\rm tr} \, \Omega^2\right)^2 - \frac{1}{2} a^2 {\rm tr}\,  \Omega^2 
- \left|\Omega a\right|^2 \, , \nonumber \\
I_3 &=& \left|\Omega^2{\bf a}\right|^2 + \frac{1}{2}  \left|\Omega{\bf a}\right|^2 {\rm tr}\,  \Omega^2 
- \frac{1}{4}a^2 {\rm tr}\,  \Omega^4 + \frac{1}{8} a^2 \left({\rm tr}\,  \Omega^2\right)^2\, , 
\end{eqnarray}
and the eigenvalues are solutions of the equations 
\be
e^6 - I_1 e^4 - I_2 e^2  - I_3 =0\, . 
\ee
The essentially new stationary motions therefore arise when $I_3\ne0$ because otherwise at least two of the eigenvalues vanish. 

\section{Brane kinematics}
\setcounter{equation}{0}

Following \cite{Russo:2008gb}, the above results for relativistic particle kinematics may be generalized to the kinematics of relativistic branes. To summarize,  we suppose that we have a $(p+1)$-dimensional surface, with local coordinates $\{\sigma^i; i=0,1,\dots p\}$,  isometrically embedded  in a $D$-dimensional Minkowski spacetime, with induced `worldvolume' metric 
\be
ds^2_{ind} = g_{ij} d\sigma^i d\sigma^j  \, , \qquad g_{ij} = \partial_i X^\mu \partial_j X^\nu \eta_{\mu\nu}\, .   
\ee
We may split $\sigma^i \to (t,\sigma^a)$ ($a=1,\dots p$) and write this induced metric as
\be
ds^2_{ind} = g_{tt} dt^2 + 2g_{t a} dt d\sigma^a  + h_{ab} d\sigma^a d\sigma^b\, . 
\ee
Its inverse can then be written as
\be
g^{ij}\partial_i\partial_j  = -u^iu^j\partial_i\partial_j  + h^{ab}\partial_a \, , 
\ee
where $h^{ab}$ is the inverse to $h_{ab}$. The vector field $u=u^i\partial_i$ is dual to the 1-form
\be
u_i d\sigma^i = - \Delta dt \, , \qquad \Delta = \sqrt{-g_{tt} + g_{ta} h^{ab} g_{tb}} \equiv {1\over  \sqrt{-g^{tt}}}\, , 
\ee
from which we see that $u^2=-1$, and hence that $u$ may be viewed as a $(p+1)$-velocity field,  which may be pushed forward to the ambient Minkowski space to give the $D$-velocity field
\be\label{braneU}
U = \left(u^i\partial_i X^\mu \right) \partial_\mu\, . 
\ee
For $p=0$ this construction yields the $D$-velocity of a particle with a worldline embedded in the $D$-dimensional Minkowski spacetime. For $p>0$ it yields the $D$-velocity field of a congruence of worldlines, one passing through
each point of the $p$-brane at fixed time $t$. We refer the reader to \cite{Russo:2008gb} for further details. 

We similarly define the brane $D$-acceleration, $D$-jerk and $D$-snap as
\be\label{braneA}
A= u^i\partial_i U \, , \qquad \Sigma = u^i\partial_i A - A^2 U\, , \qquad \Xi = u^i\partial_i \Sigma - \left(A\cdot\Sigma\right) U\, .  
\ee
Higher derivatives are defined iteratively by the brane generalization of (\ref{gar}):
\be\label{garbrane}
P_{n+1}=u^i \partial_i P_n - (A\cdot P_n)\ U\, . 
\ee

\subsection{Motion in a hot braneworld}

The stationary particle motion of subsection \ref{subsec:Acc+} may be generalized to describe a  $(D-2)$-brane,  with $(D-1)$-dimensional  Minkowski  worldvolume,   accelerating in an ambient $D$-dimensional Minkowski spacetime. 
Let $(T,X,Y, \vec Z)$ be the ambient Minkowski spacetime coordinates, with $\vec Z = (Z_1 , \dots , Z_{D-3})$,  
and consider the hypersurface,  with coordinates $(t, \vec z)$, defined by 
\be
T=T(t) \, , \qquad X= X(t) \, , \qquad Y= y \, , \qquad \vec Z = \vec z\, . 
\ee
The induced metric on the hypersurface is the Minkowski metric 
\be\label{induced}
ds^2_{ind} = -dt^2 +  dy^2 + |d\vec z|^2\, , 
\ee
provided that
\be
\partial_t T = \cosh \omega(t)\, , \qquad \partial_t X = \sinh \omega(t)\, , 
\ee
for some function $\omega(t)$. The worldvolume velocity field for this example is $u= \partial_t$, and 
hence 
\be
U = \cosh \omega \, \partial_T + \sinh\omega\, \partial_X \, . 
\ee
A calculation using (\ref{braneU}) and (\ref{braneA}) then yields
\be
A = \left(\partial_t\omega\right) \left(\sinh \omega \, \partial_T + \cosh\omega\, \partial_X \right) \, , \qquad
\Sigma = \left(\partial_t^2 \omega\right)  \left(\sinh \omega \, \partial_T + \cosh\omega\, \partial_X \right) \, , 
\ee
and so on.  In order to describe a Minkowski $(D-1)$-spacetime at Unruh temperature $T_U$ we choose
\be
\omega(t) = a t \, , \qquad a= 2\pi T_U\, . 
\ee
In this case
\be
T= a^{-1} \sinh at \, , \qquad X = a^{-1} \cosh at\, , 
\ee
and 
\be
U= \cosh\left(at\right) \, \partial_T + \sinh\left(at\right) \, \partial_X\, , \qquad
A= a \left[\sinh\left(at\right)\,  \partial_T + \cosh\left(at\right)\, \partial_X \right]\, , 
\ee
while $\Sigma=\Xi= C = \dots =0$. The acceleration is normal to the Minkowski hypersurface and has 
constant magnitude $|A|=a$. 

Now consider the hypersurface defined by
\begin{eqnarray}
T &=& a^{-1}\sinh \left[a\gamma\left(t+vy\right)\right]\, , \qquad 
X =  a^{-1} \cosh\left[a\gamma\left(t+vy\right)\right]\, ,  \nonumber\\
Y &=&  \gamma\left(y+vt\right)\, , \qquad \vec Z=\vec z \, , \qquad 
\gamma= 1/\sqrt{1-v^2}\, . 
\end{eqnarray}
The induced metric is again the Minkowski metric (\ref{induced}), as it should be since the new embedding 
differs from the old one by a boost in the $y$-direction.  However, the brane $4$-velocity is now
\be
U = \gamma\left\{ \cosh \left[a\gamma \left(t+vy\right)\right] \partial_T  + 
\sinh \left[a\gamma\left(t+vy\right)\right]\partial_X +  v \partial_Y \right\}\, , 
\ee
which has a component in the $Y$-direction expected for motion with velocity $v$ in this direction. 
The brane acceleration is 
\be
A= \gamma^2a \left\{ \sinh \left[a\gamma\left(t+vy\right)\right]\partial_T
+ \cosh \left[a\gamma\left(t+vy\right)\right] \partial_X\right\}\, . 
\ee
This is orthogonal to the brane and has constant magnitude
\be\label{gsquaredA}
|A| = \gamma^2 a\, . 
\ee
The brane jerk is
\be
\Sigma =  - v a^2\gamma^5 \left\{\partial_Y + v\cosh \left[a\gamma\left(t+vy\right)\right] \partial_T 
+ v  \sinh \left[a\gamma\left(t+vy\right)\right]\partial_X\right\}\, . 
\ee
A computation of the parameter $\lambda$ yields $\lambda=v$, so we expect a significant deviation from thermality 
unless $v\ll1$.  

A computation of the relativistic snap and crackle yields
\be
\Xi = - v^2 a^2 \gamma^4 A\, , \qquad C= - v^2a^2 \gamma^4\,  \Sigma \, , 
\ee
and from this we deduce that 
\be
P_{2k+1} = \left(-v^2 a^2 \gamma^4\right)^k A \, , \qquad P_{2k+2} = \left(-v^2 a^2 \gamma^4\right)^k \Sigma\, , 
\qquad (k=0,1,2,\dots).
\ee
and hence that  $P_{n+1}$ is of order $v^n$ for $v\ll1$.  The dimensionless parameters $\lambda^{(n)}$ defined in (\ref{params})  may be computed from these results, and one finds the
remarkably simple formula $\lambda^{(n)}=v^n$.


\section{Some Non-Stationary Motions} 
\setcounter{equation}{0}

We now consider some  non-stationary motions.   The first  is a generalization of constant proper acceleration to constant jerk, snap etc. The second case illustrates how the jerk and snap may continue to be important relative to acceleration even as a particle is brought to rest. 

\subsection{Constant $|P_n|$} 
\label{subsec:D2}

Consider a $D$-dimensional Minkowski spacetime with cartesian coordinates $(t,x,\dots)$, and a worldline with $D$-velocity
\be
U = \cosh \omega(\tau)  \, \partial_t + \sinh \omega(\tau) \, \partial_x 
\ee
where the function $\omega(\tau)$ is to be specified.  The $D$-acceleration is
\be
A =  \dot\omega\left( \sinh \omega\,  \partial_t + \cosh \omega \, \partial_x \right)\,  
\ee
where the overdot indicates differentiation with respect to proper time $\tau$.  Using the notation
$\omega^{(n)} := d^n\omega/d\tau^n$,
we find the  $D$-jerk and $D$-snap  to be given by
\be
\Sigma =  \omega^{(2)} \left( \sinh \omega\,  \partial_t + \cosh \omega \, \partial_x \right)\, , \qquad
\Xi =  \omega^{(3)} \left( \sinh \omega\,  \partial_t + \cosh \omega \, \partial_x \right)\, . 
\ee
More generally, it follows from the definition of $P_n$ that
\be
P_n = \omega^{(n)} \left( \sinh \omega\,  \partial_t + \cosh \omega \, \partial_x \right)\,  \qquad
(n>0).
\ee

The general worldline with  $P_{k+1}\equiv 0$ corresponds to the choice
\be
\omega(\tau) = p_1\tau + \frac{1}{2} p_2\tau^2 + \dots + \frac{1}{k!} p_k \tau^k\,  \qquad \left(p_k\ne0\right).
\ee
It follows that $P_k^2$ is constant on these worldlines,  and that $P_n\equiv 0$ for $n>k$.   The $k=1$ case, with $p_1=a$, yields a worldline with constant proper acceleration $|A|= a$. Integrating $dX/d\tau =U$ for this case, one finds that
\be\label{unifa}
t= a^{-1} \sinh(a \tau)\ ,\qquad x= a^{-1} \cosh(a \tau)\ , 
\ee
and hence that the worldline is one branch of the hyperbola
\be
x^2 -t^2 = a^{-2}\, . 
\ee
The asymptotes $x= \pm t$ are the Rindler horizon, and the particle remains at constant distance $a^{-1}$ from this horizon. 

In the case of constant proper jerk, one has
\be
\omega(\tau) = a\tau + \frac{1}{2} j \tau^2
\ee
where both $a$ and $j$ are constants; this choice corresponds to a worldline with time-dependent proper acceleration
$|A|=a + j\tau$, and constant proper jerk $|\Sigma |=j$.  The snap and all higher time derivatives vanish.  
Integrating $dX/d\tau =U$, we find that
\be
t(\tau)  = \int_0^\tau d\tau' \cosh \left( a\tau' + \frac{1}{2} j {\tau'}^2\right) \, , \qquad
x(\tau) =  \int_0^\tau d\tau' \sinh \left( a\tau' + \frac{1}{2} j {\tau'}^2\right) + x_0  \, , 
\ee
where $x_0$ is an integration constant; for $a\ne0$ one must choose $x_0 = a^{-1}$ in order to recover (\ref{unifa}) when $j=0$. Assuming that $a>0$, the integrals can be computed  in terms of the error function erf(x). At small $t$, one finds that the distance from Rindler horizon  is
\be
d={1\over a} + {1\over 6} j \tau ^3+{1\over 120} j a^2 \tau^5 + \dots
\ee
This distance is increasing with $\tau$ for $j>0$, and  decreasing for $j<0$ such that the particle passes through the Rindler horizon in a finite proper time. Of course, the Rindler horizon is no longer the particle's event horizon, whatever the sign of $j$, because the proper acceleration is not constant.  The actual horizon  is at
\be
x \mp t  = \frac{1}{a}  \mp \sqrt{\pi \over 2|j|}\   \exp\left({\frac{a^2}{ 2|j|}}\right) \left[ 1 
\mp {\rm erf} \left( \frac{a}{ \sqrt{2|j|}}\right) \right]\ ,
\ee
where $\mp $ signs stand for the cases $j>0$ and $j<0$ respectively.

\subsection{Welcome to Speed}

The residents of Speed, a rural town in the USA,  decide to limit the number of vehicles on the straight road passing through their town by erecting a series of roadside signs, which they must do without contravening a federal regulation that allows  only informational signs.  At  the town limit they erect a sign that says  ``Welcome to  Speed''  while another sign 50 miles down the road states ``Speed limit: 50''. If asked, the residents of Speed will state that this sign merely tells a driver that the town limit is 50 miles away, but they hope that it will be interpreted as limiting the driver's speed to 50 mph. Closer to Speed, at 30 miles from the town limit, another sign states ``Speed limit: 30''. After passing that sign, any visitor to Speed encounters an ever increasing density of signs, with the sign at  $X$ miles from the town limit stating ``Speed limit: X''.  Of course, any driver who slows down in the way intended will never get to Speed!  Here we will compute the relativistic jerk, snap and all higher kinematical quantities of a vehicle that obeys, to the letter,  the signs on the road to Speed, such that it has velocity $X$ when $X$ miles away from the town limit.  

One may choose units such that the vehicle's motion obeys the equation $dx/dt=-x$.  Equivalently,  $\dot x=-x\,  \dot t$, where the overdot indicates differentiation with respect to proper time $\tau$. Since $\dot t^2 -\dot x^2=1$, we can solve for  $(\dot t,\dot x)$ as a function of $x$, and this gives us the $2$-velocity
\be
U = \frac{1}{\sqrt{1-x^2}} \left[ \partial_t - x \partial_x\right]\, . 
\ee
This is supposed to hold only for some $x<1$ and we are interested in the limit as $x\to 0$.  One can show that
$t$ and $x$ are determined implicitly as functions of $\tau$ by
\be
e^\tau = \left(\frac{1+ \sqrt{1-x^2}}{x}\right) e^{-\sqrt{1-x^2}}\, , \qquad x= e^{-t}\, . 
\ee
Thus, $t\to\infty$ as $x\to 0$ and $t\approx \tau$ in this limit.  The $2$-acceleration is
\be
A =  \frac{x}{\left(1-x^2\right)^2} \left[-x\partial_t + \partial_x\right] \, . 
\ee
As expected this vanishes as $x\to 0$.  A computation of the jerk and snap yields
\be
\Sigma = - \frac{\left(1+ 2x^2\right)}{\left(1-x^2\right)^{\frac{3}{2}}} A\, , 
\qquad \Xi = \frac{\left(1+ 11x^2 + 6x^4\right)}{\left(1-x^2\right)^3} A\, . 
\ee
Similarly one can show that all higher  $P_n$ are non-vanishing. 
The dimensionless parameters defined in (\ref{parameters}) are
\be
\lambda = x^{-1} + 2x\, , \qquad \eta = x^{-2} + 11 + 6x^2\, . 
\ee
They are {\it never} small, and both become infinite as $x\to 0$!  Of course, both jerk and snap, and all $P_n$ with $n>0$ go to zero as $x\to 0$ but  they become {\it relatively} more important than acceleration in this limit. 

Going backwards in time, the trajectory terminates on the point $(t,x)=(0,1)$, where  the speed of light is reached. At this point,
all $|P_n|$, with $n>0$, diverge, but  $\lambda $ and $\eta $ remain finite.

\section{GEMS for free fall in black holes}

For a black hole spacetime in thermal equilibrium, a static observer at infinity perceives a heat bath at the Hawking temperature. The much higher local temperature near the horizon can be understood as an Unruh temperature resulting from the acceleration of the local frame of a static observer, but there would appear to be no similar  kinematic interpretation of the Hawking temperature  at infinity because a static observer at infinity is in free fall.  However,  if one considers an isometric global embedding  of the black-hole  spacetime in a higher-dimensional flat spacetime then one may compute the acceleration, jerk etc. with respect to the flat embedding metric.  One finds that all static observers undergo constant proper acceleration in the embedding spacetime, such that application of the Unruh formula  yields the expected local temperature required by  thermal equilibrium \cite{Deser:1998bb}. This applies, in particular, to the static observer in free fall at infinity, whose Unruh temperature is precisely the Hawking temperature. 

Within this `GEMS' approach to black hole thermodynamics, one may compute the acceleration, jerk etc. of any 
other observer, in particular of an observer in free fall towards the black hole horizon, for whom the acceleration $A$ is always orthogonal to the black hole spacetime.  If the proper acceleration is approximately constant then one expects approximate validity of the Unruh formula relating acceleration to temperature; specifically, one needs $\lambda\ll1$, where $\lambda$ is the parameter defined in  (\ref{parameters}).  We shall compute $\lambda$ for free fall from rest at infinity in Schwarzschild and Reissner-Nordstrom (RN) spacetimes, elaborating on the results of \cite{Brynjolfsson:2008uc}.  In the RN case $\lambda$ may have an  isolated zero so that it becomes necessary to consider  $\eta$ too. 

One motivation for this analysis is to see what the GEMS picture has to say about observers who fall through the horizon. One point to consider is that even if the acceleration in the flat embedding spacetime is not constant, one might still  expect a detector to detect particles whenever the acceleration is non-zero. Obviously, a detector cannot be pointlike but it may be assumed to follow the trajectory of a point particle as long as its size is small compared to the black hole. Thus, any detector of relevance to the problem is limited to detect particles with a Compton wavelength much less than $2MG$,  where $M$ is the black hole mass. This means, in particular, that the detector will not be able to detect the Hawking radiation at infinity. The equivalence principle then states  that the detector will not be able to detect  particles while in free fall through the black hole horizon. On the other hand, we can expect $|A|$ to  increase as the detector approaches the horizon,  and if  it were to increase sufficiently, so that  $|A|\gg c^4/GM$,  then a detector falling through the horizon {\it would} detect particles. There is therefore a potential conflict between the GEMS picture and the equivalence principle. As we shall see, this potential conflict does not materialize. 

Another point of interest that we briefly address is what happens to the particle after it falls through the horizon and approaches the singularity behind the horizon. 

\subsection{Schwarzschild}

The  Schwarzschild metric is
\be
ds^2 = - \left(1- 2M/r\right)dt^2 + \frac{dr^2}{1-2M/r} + r^2 \left(d\theta^2 + \sin^2\theta d\phi^2\right)\, . 
\ee
This metric can be isometrically embedded in 6D Minkowski spacetime as follows \cite{Fronsdal:1959zza}
\begin{eqnarray}\label{Sembed}
X^0 &=& \kappa^{-1} \sqrt{1-u}\,  \sinh \kappa t \, , \qquad 
X^1 =  \kappa^{-1} \sqrt{1-u}\,  \cosh \kappa t \, , \nonumber \\
X^2 &=& -\frac{1}{2\kappa} \int  \frac{du}{u^2} \sqrt{u+u^2+u^3} \\
X^3 &=& x\equiv r\cos\phi\sin\theta \, , \qquad 
X^4= y\equiv r\sin\phi\sin\theta \, , \qquad
X^5 =z \equiv r\cos\theta \, , \nonumber
\end{eqnarray}
where
\be 
\kappa = 1/4M\, , \qquad u= 2M/r\, . 
\ee
The $6$-velocity of any radial  timelike worldline has components
\begin{eqnarray}
U^0 &=& \sqrt{1-u}\,  \dot t  \cosh \kappa t  + \frac{u^2 }{\sqrt{1-u}}\,  \dot r \sinh\kappa t \nonumber\\
U^1 &=& \sqrt{1-u}\,  \dot t  \sinh \kappa t  + \frac{u^2 }{\sqrt{1-u}} \, \dot r \cosh\kappa t \nonumber\\
U^2 &=& \dot r \, \sqrt{u+u^2+u^3} \\
U^3 &=& \dot r \sin\theta\cos\phi\, , \qquad U^4 = \dot r \, \sin\theta\sin\phi\, , \qquad 
U^5 = \dot r\, \cos\theta\, .  \nonumber
\end{eqnarray}
One may verify that $U^2=-1$ using the fact that 
\be\label{timelike}
\left(1-u\right) \dot t^2 -  \frac{\dot r^2}{1-u}  =1 
\ee
on radial  timelike worldlines.  The $6$-acceleration of any radial  timelike worldline has components
\begin{eqnarray}
A^0 &=& \left[ \kappa \sqrt{1-u} \, \dot t^2 + \frac{u^2}{\sqrt{1-u}}\, \ddot r - \frac{\kappa u^3\left(4-3u\right)}{\left(1-u\right)^{3/2}}\, \dot r^2 \right]\sinh \kappa t \nonumber \\
&& \ + \left[ \sqrt{1-u} \, \ddot t + \frac{2\kappa u^2}{\sqrt{1-u}}\, \dot t\dot r \right] \cosh \kappa t 
\nonumber\\
A^1 &=& \left[ \kappa \sqrt{1-u} \, \dot t^2 + \frac{u^2}{\sqrt{1-u}}\, \ddot r - \frac{\kappa u^3\left(4-3u\right)}{\left(1-u\right)^{3/2}}\, \dot r^2 \right]\cosh \kappa t \nonumber \\
&& \ + \left[ \sqrt{1-u} \, \ddot t + \frac{2\kappa u^2}{\sqrt{1-u}}\, \dot t\dot r \right] \sinh \kappa t 
\nonumber\\
A^2 &=& \frac{\left(u+u^2+u^3\right) \ddot r - \kappa u^2\left(1+2u + 3u^2\right) 
\dot r^2}{\sqrt{u+u^2+u^3}} \\
A^3 &=& \ddot r\, \sin\theta\cos\phi \, , \qquad A^4 = \ddot r\, \sin\theta\sin\phi\, , \qquad
A^5 = \ddot r\, \cos\theta \, . \nonumber
\end{eqnarray}
One may verify using (\ref{timelike}) and its derivative, that $A\cdot U=0$. We omit the explicit expression for the 
$6$-jerk as it is rather long. 

Let us now specialize to the case of radial free fall, for which 
\be
\dot t=\frac{e_0}{(1-u)}\ ,\qquad \dot r= \sqrt{e_0^2 -1 + u}
\ee
for some constant $e_0$.  For $e_0=1$ the particle falls from rest at infinity, and this is the only case that we shall consider here.  The qualitative features are as follows. Near infinity, one has 
\be
|A|={1\over 4M} +{3\over 4 r} + {15M\over 8r^2}+  O(1/r^3)
\ee
The leading term coincides with $2\pi T_{\rm Hawking} $ as expected.
As $r$ is decreased from infinity, $|A|$ increases monotonically and at the horizon
 $|A|$ attains the value $\sqrt{7}/2M$.  The behaviour of the jerk near infinity is such that 
\be
\Sigma ^2 = {1\over 128M^3 r}+ O(1/r^2)\, . 
\ee
In fact, $\Sigma^2$ is monotonically increasing for decreasing $r$, attaining a finite value at the horizon.
The parameter $\lambda$ defined in (\ref{parameters}) has the behavior
\be\label{lamlarger}
\lambda = \sqrt{2M\over r} + O(1/r^{7/2})\, . 
\ee
It is also monotonically increasing for decreasing $r$, and at the horizon it has the value 
$\lambda=\sqrt{233/294}\cong 0.89=O(1)$. Since $\lambda $ is not small  near the horizon the  interpretation of $|A|/2\pi $ as temperature cannot be justified. Still, the presence of non-vanishing acceleration near the horizon suggests that the free falling observer could detect some sort of non-thermal radiation. We return to this point in the next section.

\subsubsection{The singularity}

In Kruskal coordinates $({\cal U},{\cal V}, \theta,\phi)$, the  Schwarzschild metric is
\be
ds^2 =  - \frac{e^{-2\kappa r}}{2\kappa^3 r} \, d{\cal U} d{\cal V} + r^2 \left(d\theta^2 + \sin^2\theta d\phi^2\right)\, , 
\ee
where $r$ is now the function of ${\cal U}$ and ${\cal V}$ defined implicitly by 
\be
{\cal U}{\cal V} = - \left(\frac{r-2M}{2M}\right) e^{2\kappa r}\, . 
\ee
The embedding in a $D=6$ Minkowski spacetime is now achieved by setting 
\be
 X^0+ X^1 =  \kappa^{-1}\sqrt{u} \, e^{-\kappa r} {\cal V}\, ,  \qquad 
 X^0 -X^1 =  \kappa^{-1}\sqrt{u} \, e^{-\kappa r} {\cal U}\, , 
 \ee
with $(X^2,X^3,X^4,X^5)$ as in (\ref{Sembed}).   Observe that not only is this induced metric non-singular at $r=2M$ but so also is the embedding. This means that we may follow the motion of the particle in the embedding space as it falls through the horizon in the black hole spacetime.  The components of the $6$-velocity on a radial timelike worldline are now 
\begin{eqnarray}
U^0 + U^1 &=& 2M \sqrt{u} e^{-\kappa r} \left[ \left(1+u^2\right) \dot {\cal V} -
 \left(1-u^2\right) \left({\cal V}/{\cal U}\right) \dot{\cal U}\right] \nonumber\\
 U^0 -  U^1 &=& 2M \sqrt{u}e^{-\kappa r} \left[ \left(1+u^2\right) \dot {\cal U} -
  \left(1-u^2\right) \left({\cal U}/{\cal V}\right) \dot{\cal V}\right] \nonumber\\
  U^2 &=& \sqrt{u+u^2+u^3} \, \dot r\\
  U^3 &=& \dot r \, \sin\theta\cos\phi\, , \qquad U^4 = \dot r \, \sin\theta\sin\phi\, , \qquad
  U^5 = \dot r\, \cos\theta\, , \nonumber
  \end{eqnarray}
 where
 \be
 \dot r = -2Me^{-2\kappa r} u \left({\cal U}\dot{\cal V} + {\cal V}\dot{\cal U}\right)\, . 
 \ee
 
As before we now focus on radial geodesics. As the singularity is approached, the acceleration and 
jerk increase without bound. The behavior of their magnitudes near the singularity is as follows
\begin{eqnarray}
| A|  &=& {3M\over r^2}+ {5\over 6 r} + {1\over 108M} + \dots \nonumber \\
| \Sigma | &=&  {9M^2\over r^4}+ {7M\over r^3} - {11\over 9 r^2} + \dots 
\end{eqnarray}
It follows that
\be
\lambda = 1+ {2r\over 9 M}- {37r^2\over 108 M^2}+  \dots
\ee
Therefore $\lambda =1$ at the singularity. This can be understood from the
fact that $\lambda^2 =1+ (\dot A)^2/A^4 $ and the second term tends to zero as the
singularity is approached. Moving away from  the singularity (and hence towards the past on any future-directed timelike geodesic) $\lambda$ first increases slightly to a maximum value and then begins to decrease
smoothly until infinity, where its  behavior is given by (\ref{lamlarger}). 

One may wonder where the particle is in the  embedding spacetime when it hits the singularity inside the horizon of the black hole metric. In other words: where in the $D=6$ Minkowski spacetime is the singularity of the Schwarzschild metric. One {\it a priori} possibility is illustrated by the example of a rigidly-rotating open string: as shown in \cite{Russo:2008gb},  the string boundary is a curvature singularity of the induced worldsheet metric but 
(as is well-known) just a null curve in the Minkowski spacetime. Could some black hole singularities be similarly interpreted as `boundaries'? This possibility is not realized by the spacelike singularity of the Schwarzschild 
black hole; which is mapped to future infinity in the $D=6$ Minkowski spacetime. In other words, a particle falling 
radially into a Schwarzschild black hole is accelerated {\it in finite proper time} to future infinity in the $D=6$ Minkowski spacetime. This is possible because the particle's acceleration in this embedding spacetime is unbounded.

\subsection{Reissner-Nordstrom}

The  Reissner-Nordstrom metric is
\be
ds^2 = - \left(1- {2M\over r}+{q^2\over r^2}\right) dt^2 + \frac{dr^2}{1-{2M\over r} +{q^2\over r^2}} + r^2 \left(d\theta^2 + \sin^2\theta d\phi^2\right)\, . 
\ee
It has two horizons at
\be
r_{\pm }= M\pm \sqrt{M^2-q^2}
\ee
The  surface gravity at $r=r_+$ is given by
\be 
\kappa = {r_+-r_-\over 2 r_+^2} . 
\ee
This metric can be isometrically embedded in a 7D  spacetime with signature $(-,+,+,+,+,+,-)$ as follows \cite{Deser:1998xb}
\begin{eqnarray}
X^0 &=& \kappa^{-1} \sqrt{1- {2M\over r}+{q^2\over r^2}}\,  \sinh \kappa t \, , \qquad 
X^1 =  \kappa^{-1} \sqrt{1- {2M\over r}+{q^2\over r^2}}\,  \cosh \kappa t \, , \nonumber \\
X^2 &=& \int dr \sqrt{ {r^2(r_+ + r_-) +r_+^2(r+r_+)\over r^2(r-r_-) }}  \\
X^6 &=& \int dr \sqrt{ {4r_+^5  r_-\over r^4(r_+ -r_-)^2} } 
\nonumber\\
X^3 &=& x\equiv r\cos\phi\sin\theta \, , \qquad 
X^4= y\equiv r\sin\phi\sin\theta \, , \qquad
X^5 =z \equiv r\cos\theta
\end{eqnarray}
A novel feature of this  case is that  the flat embedding spacetime has  {\it two} time dimensions.  In this case the requirement that $P_n $ be timelike  for $n>0$ should be replaced by the requirement that $P_n $  be orthogonal to $U$ for $n>0$. The two conditions are equivalent  in Minkowski spacetime, but it is the latter that should be applied when there is more than one time dimension. Thus understood, our definitions for jerk, snap etc. continue to be the unique solution to the other requirement  that $P_n\equiv 0$ implies $P_{n+1}\equiv 0$, but it should now be appreciated that any of the $P_n$ with $n>0$ may be timelike or null as well as spacelike. 

The free fall trajectory is now given by
\be
\dot t={e_0\over 1-{2M\over r}+{q^2\over r^2} }\ ,\qquad \dot r= \sqrt{e_0^2-(1-{2M\over r} +{q^2\over r^2})}
\ee
Using these expressions, we can compute the 7-velocity, and then the acceleration and jerk. 
We again consider only the case for which $e_0=1$, corresponding to a particle falling freely from rest at infinity.
In brief, the temperature agrees with Hawking's at infinity, as expected. At finite values of $r$ (but not too close to the horizon) one finds that 
\bea
|A| &=& \kappa \left[1 +{3M\over r} +{3\over 2}(5M^2-q^2){1\over r^2} +O({1\over r^3} )\right] \nonumber \\
|\Sigma| &=& \kappa^2\ \sqrt{2M\over r}\left[1 +{24M^2-q^2\over 4M\ r} +O({1\over r^2})\right] \nonumber\\
|\Xi| &=& \kappa^3\ {2M\over r}\left[1 +{18M^2-q^2\over 2M\ r} +O({1\over r^2})\right]\, , 
\eea
and hence
\be
\lambda = \sqrt{2M\over r} \left[1 - {q^2\over 4M\ r} +O({1\over r^2})\right]\, , \qquad
\eta =  {2M\over r} - {q^2\over  r^2} +O({1\over r^3})\, . 
\ee
Some interesting features appear near the horizon. For the extremal RN black hole, with $q=M$, it was pointed out in \cite{Brynjolfsson:2008uc} that $A^2<0$ near the horizon because the motion  is geodesic motion on $adS_2$, for which $A^2=-R^{-2}$,  where $R\sim 1/M$ is the adS radius. In fact, our calculations show that  $A^2$ becomes negative near the horizon  when  $q/M>(q/M)_1\approx 0.8$ (otherwise remaining positive everywhere). In addition, $\Sigma^2$ becomes negative near the horizon when  $q/M> (q/M)_2 \approx 0.62$. It follows that for $(q/M)_2< q/M < (q/M)_1$ there is a point on the particle's worldline at which $A^2>0$ and $\Sigma^2=0$, so that $\lambda=0$.  However, the value of $A^2$ at this point does {\it not} correspond (via the Unruh formula) to the local temperature of the black hole. This is because the snap is not negligible at this point;  in fact, $\eta >1$. This example thus shows that $\lambda\ll1$ is not sufficient  for an application of the Unruh formula.

\section{Discussion}

In this paper we have presented a completion of relativistic kinematics for particle motion in a $D$-dimensional Minkowski spacetime, taking into account not just  the particle's relativistic $D$-velocity $U$ and $D$-acceleration 
$A$  that one can find discussed in any relativity textbook, but also the  particle's  $D$-jerk, $D$-snap, and all higher time derivatives of the acceleration.  In the instantaneous rest-frame one can define the proper jerk {\bf j}  as in non-relativistic mechanics,
but the $D$-vector jerk should not be  defined  simply as the proper-time derivative of the $D$-acceleration
because (i) it is  not orthogonal to the $D$-velocity $U$, and hence may be timelike, and (ii) it does not  vanish for worldlines of constant  $|A|$. We showed in a previous paper \cite{Russo:2008gb} that one can define a relativistic jerk $\Sigma$ that is both orthogonal to $U$ and zero on worldlines of constant $|A|$. In fact,  these conditions essentially determine $\Sigma$, up to a scale which is fixed by requiring  $\Sigma = {\bf j}\cdot \partial_{\bf x}$   in the instantaneous rest-frame.

Here, we similarly defined a relativistic snap $\Xi$ that is orthogonal to $U$ and vanishes on worldlines of constant proper acceleration. These conditions essentially determine $\Xi$ but one finds that   $\Xi \ne {\bf s}\cdot \partial_{\bf x}$ in the instantaneous rest-frame, where ${\bf s}= d{\bf j}/dt$.  In contrast to the notions of proper acceleration and proper jerk, there is an ambiguity in the definition of proper snap, arising from the fact that the triple derivative with respect to coordinate time $t$ does not coincide in the instantaneous rest-frame with the triple derivative with respect to proper time, whereas there is a coincidence for single and double derivatives. This ambiguity could be exploited to define the proper snap to be the spatial vector that one gets from $\Xi$ in the instantaneous rest-frame. In any case, the definition of $\Xi$ is unambiguous given the stated conditions, and we may then define an infinite sequence of spacelike $D$-vectors $P_n$, with $P_1=A$, $P_2=\Sigma$ and  $P_3=\Xi$, such that  $P_n\equiv 0$ implies $P_{n+1}\equiv 0$. 

For simplicity of presentation, we concentrated on the case of particle motion in Minkowski spacetime, but there is a natural generalization to motion in an arbitrary curved spacetime obtained by the replacement of the time-derivative by a covariant time derivative. Thus
\be
A=  \frac{DU}{d\tau}\, , \qquad  P_{n+1} = \frac{DP_n}{d\tau} - \left({A\cdot P_n} \right)U \qquad (n=1,2,\dots)
\ee
where, for any $D$-vector $V$,
\be
\left(DV\right)^\mu = dV^\mu + U^\lambda \Gamma^\mu_{\lambda\nu} V^\nu
\ee
where $\Gamma$ is the standard affine connection. As in the Minkowski case, $P_n\equiv 0$ implies $P_{n+1}\equiv 0$, so geodesics have the property that $P_n=0$ for all $n>0$. 

We used our results on relativistic kinematics to present a classification of stationary motions. These motions have the feature that 
the $D$-vectors $P_n$ ($n>0$) are also all vectors in a real vector space ${\cal V}$, generically  of dimension $(D-1)$, spanned by 
$(P_1,\dots, P_{D-1})$. In addition the scalar $P_n^2$ is constant for all $n$. In special cases, ${\cal V}$ has dimension $k<(D-1)$ 
and basis $(P_1,\dots, P_k)$.  We gave complete results for $D=4$, which agree with those found in  \cite{Letaw:1980yv} using a Frenet-Serret type  analysis, and we deduced some general features that apply  for all $D$. 

A case of particular interest, actually two cases in the classification of \cite{Letaw:1980yv},  is that of constant proper acceleration with a constant velocity `drift' in an orthogonal  direction. For zero drift velocity this reduces to the much-studied case of constant proper acceleration, for which the relativistic jerk $\Sigma$ is identically zero; this was classified as a separate case in \cite{Letaw:1980yv}. For non-zero drift velocity,  $\Sigma$ is non-zero and the 
proper acceleration is no longer constant.  We presented a brane generalization of these cases, defining a brane 
to be stationary when, roughly speaking, all points on it move on stationary worldlines.  Specifically, we considered a relativistic $(D-2)$-brane, with a $(D-1)$-dimensional Minkowski worldvolume,  undergoing constant uniform acceleration  in an orthogonal direction, following the definitions of our previous paper \cite{Russo:2008gb}. 
There is an ambiguity in what is meant by the `brane' in this example, because the worldvolume can be foliated by flat $(D-2)$ spaces  in different ways that are related by a worldvolume  Lorentz boost. In a boosted frame, each point on the brane drifts with constant velocity in addition to its constant orthogonal acceleration. For the brane, there is no physically significant distinction between constant proper acceleration and the stationary motion that we call acceleration with drift.  

Of course, the distinction between pure acceleration and acceleration with drift becomes physically significant when one considers a brane with a `marked point', which one could interpret as the location of a particle detector. In the case of zero drift velocity, one expects the detector to behave as if it is immersed in a heat bath at the Unruh temperature associated to the orthogonal acceleration. In the case of non-zero drift velocity, one expects the detector to behave as if it is in uniform motion in a heat bath, so our brane example provides a new context in which to 
consider this problem. The problem itself requires consideration of quantum field theory, so we defer a full analysis to a separate paper. 

We also presented various examples of non-stationary motion. Given the importance of constant proper acceleration, it is natural to consider motions of constant proper jerk, or constant proper snap.  We also described a non-stationary motion in which a particle is brought to rest in such a way that the jerk and snap become more and more important, relative to acceleration, as the point of rest is approached. 

Another interesting class of examples is provided by the GEMS approach to black hole  thermodynamics. The black hole spacetime is viewed, for kinematical purposes,  as the $(1+3)$-dimensional worldvolume of  a $3$-brane globally embedded in a higher-dimensional flat spacetime, which may be Minkowski but generically  has extra time dimensions too.  Motions  `on the brane'  then lift to motions in the flat embedding spacetime.  In the case of static black holes, a static observer undergoes constant proper acceleration in the embedding space, with a magnitude such that the Unruh temperature coincides with the local temperature required for thermal equilibrium  in the presence of Hawking radiation  \cite{Deser:1998bb}.  Here we have considered, as in \cite{Brynjolfsson:2008uc},  the motion in the embedding spacetime of an observer  who is in radial free fall towards the black hole horizon.  It is often stated that such an observer will cross the event horizon without noticing the Hawking radiation that would be noticed by a static observer near the horizon.  Does the  GEMS picture support this statement? 

In an attempt to answer this question, we computed the acceleration $A$,  jerk $\Sigma$ and snap $\Xi$ 
for  a particle that falls radially from rest into a black hole, Schwarzschild or Reissner-Nordstrom, of mass $M$.    Although the proper acceleration is constant only at infinity, such that $|A|= 2\pi T_{\rm Hawking}$, this state of affairs  changes continuously, and initially very slowly.  For a Schwarzschild black hole the magnitude of the infalling particle's acceleration increases monotonically, but it remains finite and its value at the horizon  is $5.3$ times as large as its value at infinity.  Either $\Sigma$ and/or $\Xi$ become large near the horizon, as measured by the dimensionless parameters $\lambda= |\Sigma|/A^2$ and $\eta= |\Xi|/|A|^3$, so the Unruh formula is inapplicable in the sense that we do not expect thermality. Nevertheless,  the fact that $A$ is non-zero suggests that a particle detector would detect particles, even as it falls through the horizon. Potentially, this could violate the equivalence principle if $A$ were large enough because a freely-falling observer should not be able to detect by local measurements whether (s)he is near a black hole or in empty space. Fortunately, the fact that  $|A|\sim M^{-1}$ means that  the typical wavelength of the particles that might be detected is as large as the black hole, so even if it could be detected the process of detection could not be interpreted as a {\it local} measurement.  

Finally,  it seems possible that the general formalism of relativistic kinematics described here will find  applications in an astrophysical context, such as shock waves in relativistic media or radiation from cosmic strings. As we mentioned in the introduction,  relativistic jerk is certainly relevant to the problem of radiation reaction, and gravitational radiation reaction must be taken into account when considering processes such as black hole collisions. 

\bigskip
\noindent
\section*{Acknowledgments}

We thank Gary Gibbons for pointing out the relevance of the Lorentz-Dirac equation and  Thomas Heinzl for
bringing to our attention the paper  \cite{Taub:1948zza} of Taub. We thank an anonymous referee for noticing
various typographical errors and a confusion of notation in an earlier version. 
JGR acknowledges support by MCYT FPA 2007-66665.
PKT is supported by an ESPRC Senior Research Fellowship, and he thanks the University of 
Barcelona for hospitality.

\setcounter{section}{0}

\end{document}